# Transitions between distinct compaction regimes in complexes of multivalent cationic lipids and DNA

Oded Farago[1], Kai Ewert[2], Ayesha Ahmad[2], Heather M. Evans[2], Niels Grønbech-Jensen[3], and Cyrus R. Safinya[2]

[1]Department of Biomedical Engineering, Ben Gurion University, Be'er Sheva 84105, Israel
[2]Materials Department, Physics Department, and Biomolecular Science and Engineering Program, University of California, Santa Barbara, CA 93106.
[3]Department of Applied Science, University of California, Davis, CA 95616.

**Abstract**

Cationic lipids (CLs) have found widespread use as nonviral gene carriers (vectors), including applications in clinical trials of gene therapy. However, their observed transfection efficiencies (TEs) are inferior to those of viral vectors, providing a strong incentive for a detailed understanding of CL–DNA complex behavior. In recent systematic studies employing monovalent as well as newly synthesized multivalent lipids (MVLs), the membrane cationic charge density has been identified as a key parameter governing the TE of lamellar CL–DNA complexes. In this work, we use X-ray scattering and molecular simulations to investigate the structural properties of complexes containing MVLs. At low mole fraction of neutral lipids (NLs), $\Phi_{NL}$, the complexes show dramatic DNA compaction, down to essentially close packed DNA arrays with a DNA interaxial spacing $d_{DNA} = 25$ Å. A gradual increase in $\Phi_{NL}$ does not lead to a continuous increase in $d_{DNA}$ as observed for DNA complexes of monovalent CLs. Instead, distinct spacing regimes exist, with sharp transitions between them. Three packing states have been identified: (i) close packed, (ii) condensed, but not close packed, with $d_{DNA} = 27$–$28$ Å, and (iii) an expanded state, where $d_{DNA}$ increases gradually with $\Phi_{NL}$. Based on our experimental and computational results, we conclude that the DNA condensation is mediated by the multivalent cationic lipids, which assemble between the negatively charged DNA rods. Quite remarkably, the computational results show that the less tightly packed structure in regime (ii) is thermodynamically more stable than the close packed structure in regime (i). Accordingly, the constant DNA spacing observed in regime (ii) is attributed to lateral phase coexistence between this stable CL–DNA complex and neutral membranes. This finding may explain the reduced TE measured for such complexes: Transfection involves endosomal escape and disassembly of the complex, and these processes are inhibited by the high thermodynamic stability. Our results, which demonstrate the existence of an inverse correlation between the stability and transfection activity of lamellar CL–DNA complexes are, therefore, consistent with a recently proposed model of cellular entry.

**Introduction**

Somatic gene therapy holds great promise for future medical applications including, for example, novel treatments for various inherited diseases and cancers (1, 2). Complexes composed of cationic lipids (CLs) and DNA, named lipoplexes, constitute one of the most promising nonviral gene delivery systems. They have attracted considerable attention due to their inherent advantages over viral delivery methods (3–5). These advantages include simple and variable preparation, unlimited length of the transported DNA, and lack of a specific immune response due to the absence of viral peptides and proteins (5–8). However, their gene transfer efficiency is currently considerably lower than that of viral vectors and substantial improvements are required before CL–DNA complexes will be viable for therapeutic purposes.

The improvement of lipid vectors requires a better understanding of their mechanism of transfection, and the chemical and physical parameters of CL–DNA complexes that influence it. Condensation of negatively charged DNA into CL–DNA complexes is thought to aid the delivery of the DNA to the cell nucleus by protecting it from enzymatic degradation, facilitating binding to the negatively charged plasma membrane, and aiding penetration of the DNA into the cytosol (9). The mechanism of transfection and the transfection efficiency (TE; the ability to transfer foreign DNA into a cell followed by expression) is strongly influenced by lipoplex properties such as their supramolecular structures, membrane charge density ($\sigma_M$) and lipid/DNA charge ratio. The two predominant phases of CL–DNA complexes are (a) a multilamellar phase where DNA monolayers are intercalated between lipid bilayers ($L_\alpha^C$) (10); and (b) the inverted hexagonal phase with DNA encapsulated within cationic lipid monolayers tubes and arranged on a two-dimensional hexagonal lattice ($H_{II}^C$) (11). Recent systematic studies employing a series of newly synthesized multivalent cationic lipids (MVLs) with headgroup charges varying from +2 e to +5 e (12, 13) have shown that the TE of lamellar complexes follows a universal, bell-shaped curve as a function of $\sigma_M$ (12, 14). At optimal (intermediate) charge densities, their TE equals that of complexes in the $H_{II}^C$ phase (12). In contrast to $L_\alpha^C$ complexes, the TE of $H_{II}^C$ complexes is independent of $\sigma_M$ (8, 15). Based on these and other observations, a model of cell entry has been proposed that describes transfection as a process involving adsorption and entry (via endocytosis) of CL-DNA complexes into the cell, followed by the release of the DNA into the cytoplasm and its delivery to the nucleus (8, 16, 17). This model suggests that in the low charge density regime, the TE of $L_\alpha^C$ complexes increases with $\sigma_M$ due to improved endosomal escape (12, 14). Recent computer simulations of lamellar complexes with monovalent lipids support this viewpoint by showing that increasing $\sigma_M$ reduces the mechanical stability of the complex, thus making it more susceptible to pore formation and disintegration (18). In the high $\sigma_M$ regime, the TE of lamellar DNA complexes of MVLs may be limited by the ability of the negatively charged DNA to dissociate from the cationic lipids. The packing of DNA by MVLs should, therefore, be closely investigated and new condensation strategies that facilitate more efficient DNA release from lipoplexes must be considered.

The DNA interaxial spacing, $d_{DNA}$, serves as a measure of the degree of DNA compaction in lamellar complexes. Here, we report on a combination of experimental and computational studies examining the dependence of $d_{DNA}$ on the mole fraction of MVLs (i.e., the ratio of MVLs to neutral lipids) using X-ray scattering and molecular simulations. In previous studies of complexes containing mixtures of the monovalent CL DOTAP and the neutral lipid (NL) DOPC, we have found that the DNA spacing increases *continuously* as DOPC is added to the system (10, 19). Thus, the fraction of NL determines the degree of compaction, indicated by the DNA interaxial distance $d_{DNA}$. This distance has been found to vary from a minimal value of $d_{DNA} \approx 25$ Å (tight packing), determined by the size of the (hydrated) DNA rod for systems without DOPC, to $d_{DNA} \approx 70$ Å (loose packing) for a high content of NLs. In contrast, our X-ray diffraction data for complexes containing MVLs reveal a markedly different behavior. Instead of a continuous increase in $d_{DNA}$, we find that the DNA spacing changes in a *discrete*, step-like, fashion between three distinct packing regimes: At high fractions of MVLs, the DNA molecules are tightly packed and their spacing does not increase upon adding DOPC but rather appears level and "locked" ("Plateau I"). As more DOPC is added, a slight but sudden increase in $d_{DNA}$ is observed. This change in $d_{DNA}$ marks the transition to the second regime where, again, over a range of DOPC concentrations, the DNA spacing remains constant. The DNA spacing in this regime ("Plateau II") is typically 2–3 Å larger than its minimal value at close packing in the Plateau I regime. Finally, at very large concentrations of DOPC, the DNA molecules "break free" and their spacing increases gradually with increasing amounts of DOPC, as in the case of monovalent CLs. The transition to this third concentration regime is very sharp (increasing the DOPC mole fraction by a few percent results in a $d_{DNA}$ increase of several Angstroms) and appears to be discontinuous.

We use computer simulations to gain insight into the nature of these distinct packing regimes and the transitions between them. The simulations utilize a molecular model where both the lipids and DNA rods are represented by coarse-grained (CG) structures (i.e., *not* in a fully atomistic representation) consisting of a relatively small number of "particles" interacting with each other via steric, hydrophobic, and electrostatic interactions. This simplified CG representation reduces computational resource requirements and, thus, enables simulations of mesoscopically large complexes over experimental timescales. It also helps to identify the non-specific interactions and mechanisms that govern the mesoscopic statistical–mechanical behavior of the system. A similar computational approach has been previously used to reproduce and analyze the experimental data for CL–DNA complexes with monovalent CLs (18, 20). As in our previous computational study, the simulation results reported here for MVL–DNA complexes are in very good agreement with the experimental data. Most significantly, we show that the tight packing of DNA rods is driven by the presence of MVLs that bind adjacent DNA rods to each other, overcoming the electrostatic repulsion of the negatively charged DNA rods. Like-charge attraction mediated by multivalent *counterions* has been widely observed and analyzed in studies of bulk (3D) biopolymer systems such as DNA, actin, and microtubules (21–25). DNA condensation has also been reported in 2D for DNA within CL–DNA complexes in the presence of di- and trivalent cations (26), but not as a result of complexation with multivalent *lipids*. Starting from the most condensed structure (Plateau I; lowest content of NLs), our simulations show that the addition of NLs to the complex only leads to a slight increase in the DNA interaxial spacing. Therefore, the area density of the lipids increases. Eventually it reaches a critical value, at which the complexed membranes undergo a transition from the fluid state to a gel-like state. The term "gel" (which is not always well-defined and sometimes confused with "glass") is herein defined as referring to a state in which (i) no diffusion of the lipids is observed on the timescale of the simulations, and (ii) the lipids exhibit local (short-range) hexagonal order. At the fluid-gel transition point, the increase of $d_{DNA}$ with $\Phi_{NL}$, the mole fraction of NLs, stops (Plateau II). This occurs because the assembly of DNA with the membranes in the gel state is energetically so favorable that NLs added into the complex laterally phase separate in order not to disrupt this highly stable structure. This lateral phase coexistence persists over a range of $\Phi_{NL}$, until coalescence into a single phase of (fluid) complexed membranes becomes thermodynamically feasible. The final phase transition into a single-phase state is characterized by an abrupt increase in $d_{DNA}$ which, upon further addition of NLs, continues to grow continuously ("Unlocked Regime"). The DNA rods are now spatially well separated and the multivalent cationic headgroups can no longer bind them to each other, resulting in overall repulsive DNA–DNA interactions.

Examining our published TE data in the light of these results supports and adds to its previous interpretation (12). High $\sigma_M$ complexes (Plateau I) exhibit suboptimal TE, which has been attributed to the inability of the DNA to dissociate from the highly charged membranes and become available for expression. In this regime, TE improves with increasing DOPC content in the membrane and the weakening of the binding forces. Beyond an optimal TE, an exponential decrease over several orders of magnitude in TE is observed (12). The transfection efficiency starts to fall off at a range of concentrations that corresponds to the second plateau in the $d_{DNA}$ vs. $\Phi_{DOPC}$ plots, i.e., the region of coexistence between complexed membranes in the gel state and free fluid membranes of NL. As argued above, the assembly of DNA arrays with membranes in the gel state containing cationic MVLs forms a very stable structure which, quite expectedly, exhibits reduced TE.

## Results and Discussion

**X-ray diffraction.** The chemical structures of the lipids used in this study are shown in Fig. 1. Monovalent 2,3-dioleyloxy-propyl-trimetylammonium chloride (DOTAP) and neutral 1,2-dioleoyl-*sn*-glycerophosphatidylcholine (DOPC) are commercially available, while the two multivalent cationic lipids (MVL3 and MVL5) were synthesized in our lab (12,13). MVL3 and MVL5 carry 3 and 5 charges, respectively, in the fully protonated state. The structure of the complexes formed by combining DNA and cationic liposomes consisting of mixtures of DOPC and MVL3 or MVL5 at different ratios has been determined by using small angle X-ray scattering (SAXS). Fig. 2A shows typical X-ray diffraction patterns of CL–DNA complexes containing varied amounts of MVL (MVL3 (left) or MVL5 (right)) and DOPC. These samples have been prepared at a charge ratio (CL/DNA) of 2.8 in the presence of Dulbecco's modified Eagle's medium (DMEM), the cell culture medium used in our transfection studies. DNA complexes of both MVLs, at all investigated CL/NL ratios, form the lamellar ($L_\alpha^C$) phase. The sharp peaks, labeled $q_{001}$, $q_{002}$, $q_{003}$, respectively, arise from the lamellar structure and give the membrane spacing d consisting of the lipid bilayer thickness ($\delta_L$) plus the thickness of the water/DNA layer between them ($\delta_W$): $d = \delta_L + \delta_W = 2\pi/q_{001}$. The diffuse weaker peak, labeled $q_{DNA}$, results from one-dimensional ordering of the DNA molecules sandwiched between the lipid bilayers and gives $d_{DNA}=2\pi/q_{DNA}$. The membrane spacing d does not change much and remains $d \approx 66$–$69$ Å for both types of MVLs over the whole range of MVL/NL ratios. The scans also show that differences in the DNA spacing between complexes of different MVLs are only seen at high concentrations of NL (scans on white background).

To assess the packing of the DNA arrays more quantitatively, $d_{DNA}$ is plotted as a function of decreasing mol% DOPC ($\Phi_{DOPC}$) in Fig. 2B. Results are shown for experiments performed in DMEM at a CL/DNA charge ratio of 2.8 (solid symbols), as well as in water at the isoelectric point, CL/DNA=1, (open circles). The latter experiments produce results that are similar but shifted to higher $\Phi_{DOPC}$. Unlike the semi-continuous dependence of $d_{DNA}$ on $\Phi_{DOPC}$ observed for complexes of monovalent DOTAP (see below), the DNA spacing in MVL-containing complexes seems to change in discrete steps. The distance between the DNA rods appears to be "locked", first at around 25 Å, and then at around 27 Å for MVL5 and 28 Å for MVL3. Upon further increase of the amount of DOPC in the membrane, the DNA rods "unlock" and $d_{DNA}$ increases with $\Phi_{DOPC}$.

For comparison, Fig. 2C shows the DNA spacing as a function of $\Phi_{DOPC}$ for lamellar complexes containing mixtures of DOPC with the monovalent lipid DOTAP. The figure shows results for complexes at CL/DNA=2.8 in DMEM (closed circles), isoelectric complexes in water (open circles), as well as the theoretical prediction for the latter based on a simple packing argument (dashed line) (19). For large $\Phi_{DOPC}$, the repulsive DNA–DNA Coulombic interactions force the DNA rods to occupy the entire membrane area, therefore leading to the scaling behavior $d_{DNA} \sim (\Phi_{DOTAP})^{-1} = (1-\Phi_{DOPC})^{-1}$ (18–20). In isoelectric systems, the DNA spacing at small $\Phi_{DOPC}$ levels off above the values of $d_{DNA}$ predicted by the packing argument. In this regime, $d_{DNA}$ is constant and determined by the hydrated diameter of the DNA rods. Consequently, the lipid area density begins to fall below its equilibrium value with the decrease in $\Phi_{DOPC}$, effectively inducing attractive "elastic" forces in the membranes that act against the repulsive Coulombic and short-range steric (hard core and hydration) forces (18, 20).

For the MVL-containing complexes, the DNA spacing at tight packing (Plateau I) is also dictated by the hydrated size of the DNA rods. As in the case of complexes with DOTAP, the steric repulsive forces between the DNA rods are balanced by the attractive "elastic forces" which are induced in the complexed membranes. In addition, however, attractive electrostatic forces mediated by the presence of multivalent headgroups are present. The addition of DOPC (increase of $\Phi_{DOPC}$) imposes a strain on this closely-packed configuration, until adjacent DNA rods can no longer be held at a fixed (minimal) separation. At very high mol% of DOPC, a transition from a condensed to an expanded state (Unlocked Regime) is observed. In this regime, the DNA rods are well separated ($d_{DNA}>30$ Å) and their spacing increases with $\Phi_{DOPC}$. As in monovalent systems, the electrostatic DNA–DNA interactions in this expanded state are repulsive. Between Plateau I and the Unlocked Regime, the complex exists in another, slightly less condensed state in which the DNA spacing attains a constant value of about 27–28 Å (Plateau II). Our simulations results, presented in the next section, suggest that the second plateau is associated with the formation of a highly stable complex structure which is not broken up by addition of DOPC to the membrane.

**Computer Simulations.** We have used computer simulations to obtain insight into the structure and properties of CL–DNA complexes containing MVLs. Our model employs a coarse grained (CG) description of both the lipids (modeled as trimers consisting of one hydrophilic and two hydrophobic beads (27, 28)) and the DNA rods (represented as rigid cylinders carrying uniform charge density), *without* the explicit presence of solvent (but assuming that electrostatic interactions take place in a medium with relative dielectric constant ε=78). A similar model has been successfully used to reproduce and analyze the behavior of DOTAP/DOPC–DNA complexes (18, 20). In the DOTAP model, the charge of the lipid was located at the center of the hydrophilic bead. For the MVL5 model used here, we assume that the lipid carries a single +5 charge rather than five point charges of size +1, which reduces the computational overhead associated with the calculation of the electrostatic interactions. Three models have been tested for MVL5 (see Fig. 3), in which the multivalent +5 charge is: (i) located at the center of the hydrophilic bead, (ii) located at the tip of the molecule, and (iii) connected to the tip of the molecule by an inextensible tether (i.e., a tether that cannot be stretched beyond a certain maximal length but has no energy at all permitted distances) of varying length $l$, $0 < l < 1.5b$, where $b \approx 6.3$ Å is the diameter of the beads. The point charge is not allowed to penetrate the beads or the DNA rods. Only in the third model, the position of the charge with respect to the lipid is allowed to change with time (through the variations in the length of the tether). Therefore, this model best mimics the effective cationic charge distribution around the hydrophilic head group of MVL5 by smearing the cationic charge distribution. Our choice of maximal tether extension ($l = 1.5b \approx 1$ nm) reflects the approximate position of the charge furthest away from the hydrophobic part as estimated using molecular models. The average extension measured in the simulation is $l \approx b$, which is roughly where the majority of the charges are located. Notably, only in model (iii) can the multivalent charge get into the small spaces between the closely packed DNA rods and be positioned in the vicinity of the DNA surfaces. The fact that this model exhibits DNA condensation while the others do not lends support to the notion that the multivalent charge of the lipids induces attractive DNA–DNA interaction at short range. This is consistent with previous studies where the short-range nature of DNA condensation by multivalent *counterions* has been demonstrated (21).

We have simulated isoelectric complexes without added counterions and under conditions of vanishing surface tension. The mole fraction of MVLs has been varied systematically by adding an appropriate number of NLs. In the simulations, we have measured the mole fraction of multivalent lipid ($\Phi_{MVL}(x)$) and the local area per lipid ($a(x)$) as a function of the position within a unit cell of the complex, as well as the DNA spacing ($d_{DNA}$). The parameter x, $0 \leq x < d_{DNA}$, spans the interval between adjacent DNA rods, with x = 0 and x = $d_{DNA}$ corresponding to lipids located right above or below the DNA rods, while x = $d_{DNA}/2$ corresponds to the middle of the unit cell (halfway between adjacent DNA rods). Because the uptake of NLs by the complex is governed by its thermodynamic stability, we have attempted to evaluate the free energy F of the complex. This has been done by measuring the energy of complexes consisting of 10 equally spaced DNA rods, and making the assumption that the dominant entropic contribution is due to the mixing of charged and neutral lipids, which may be evaluated by

$$-TS \approx k_B T (L_x / d_{DNA}) L_y \int_0^{d_{DNA}} a(x)^{-1} [\Phi_{MVL}(x) \ln \Phi_{MVL}(x) + \Phi_{NL}(x) \ln \Phi_{NL}(x)] dx, \quad (1)$$

where $L_x$ and $L_y$ denote, respectively, the lengths of the complex perpendicular and parallel to the DNA axis, $k_B$ is the Boltzmann constant, T is the temperature, and $\Phi_{NL}(x) \equiv 1-\Phi_{MVL}(x)$ is the local mole fraction of NLs. In the following, we use $\Phi_{MVL}$, $\Phi_{NL} = 1-\Phi_{MVL}$ and $a$, to denote the *mean* mole fractions of multivalent and neutral lipids and area per lipid, which are distinct from the corresponding *local* quantities defined earlier in this paragraph. The entropic contribution has been included in the free energy calculation only for complexes with fluid membranes where the lipids are mobile and can mix (see discussion below about complexes with membranes *not* in the fluid phase). The free energy gain per neutral lipid added to the complex, $\mu_{complex}(\Phi_{NL})$, is compared with the chemical potential of lipids in free (non-complexed) neutral membranes, $\mu_{free}$, to determine whether the uptake of NLs by the complex proceeds or is disfavored over the formation of a coexisting phase of neutral membranes. In previous studies we have found $\mu_{free} = -10.31\ k_BT$ (28). Fig. 4 shows the energy difference per unit length of one unit cell of the complex, $\Delta$, calculated as

$$\Delta \equiv \frac{F - \mu_{free} N}{L_y (L_x / d_{DNA})}, \qquad (2)$$

where $N \geq 0$ denotes the number of neutral lipids added to some reference state of the complex, for which we define $F = 0$. As shown in Fig. 4, upon adding NLs to the complex and, hence, increasing $\Phi_{NL}$, the excess free energy $\Delta$ monotonically decreases. This means that uptake of NLs by the complex is thermodynamically more favorable than the formation of a coexisting phase of neutral membranes. At a concentration $\Phi_{NL} = \Phi_1$ of NLs, however, $\Delta$ sharply increases. This marks the onset of phase separation. Note that this phase separation may be macroscopic, i.e., between the complexes and the neutral membrane or a lateral phase separation within the supramolecular assembly. The latter scenario, supported by microscopy and transfection data, is schematically depicted in Fig. 5C. The raw data show that the dramatic increase in $\Delta$ is not only due to the second term in the numerator on the right hand side of Eq. 2. In fact, the free energy F of the complex itself increases. We thus conclude that at the concentration $\Phi_1$, the complex reaches a particularly stable configuration which resists further dilution of the MVLs in the complexed membranes. Only at a higher concentration $\Phi_2$ does the inclusion of NLs in the complexed membranes become thermodynamically favorable again. Thus, over the range $\Phi_1 < \Phi_{NL} < \Phi_2$, the addition of NLs to the system does not affect the DNA spacing.

With the above conclusion in mind, we can now plot the computed DNA spacing as a function of $\Phi_{NL}$. The results, given in Fig. 5A, show the existence of two distinct $d_{DNA}$ plateaus. In the first, $d_{DNA}$ achieves the minimum possible value determined by the physical size of the DNA (close packing), which in our model was set to $\approx 20$ Å, the bare diameter of DNA. The experimental value is $d_{DNA} \approx 24$ Å, corresponding to the size of a DNA rod with a hydration shell around it. The second plateau in the plot (Plateau II) represents the region of phase coexistence. The DNA spacing in the Plateau II regime is 2–3 Å larger than in the close-packed CL–DNA complex (Plateau I), in agreement with the experimental data. For large values of $\Phi_{NL}$, a first order transition back into a single phase takes place, appearing as a sharp discontinuity in the value of $d_{DNA}$. Thus, our computational results reproduce the unexpected experimental observation of the existence of two plateau regions associated with distinct condensed structures. The numerical results have to be compared with the corresponding experimental data for isoelectric MVL5 complexes prepared without salt (open symbols in Fig. 2B). Here, we have a spectacular agreement between theory and experiments with regards to the concentration range of the phase coexistence region ($\Phi_{NL} \approx$ 80–85%), and the difference in $d_{DNA}$ between the two plateaus ($d_{DNA}^{Plateau\ II} - d_{DNA}^{Plateau\ I}$ = 2–3 Å). The numerical value of the distance of closest approach between adjacent DNA rods is different from the experimental one, but this is merely a matter of definition of the DNA diameter in the simulations.

Our simulations show that the condensed and expanded CL–DNA complexes differ in the distribution of charges in their membranes. Fig. 5B shows $\Phi_{MVL}(x)$, the mole fraction of MVL as a function of the position within a unit cell of the complex for several values of the *mean* mole fraction $\Phi_{MVL}$. In the condensed structures ($\Phi_{MVL}$ = 0.2, 0.171), the MVLs tend to accumulate in the small regions midway between adjacent DNA rods. This is, indeed, the position where one expects to find them if they were to serve as condensation (bridging) agents for the DNA rods. In the expanded state ($\Phi_{MVL}$=0.133), the middle of the unit cell (x=0.5) contains almost exclusively NLs, while the MVLs migrate towards the edges. This charge distribution resembles, albeit in a more polarized fashion (due to the multivalent nature of the charged lipids), the distribution of NLs and CLs in complexes of monovalent lipis (20).

The different structures formed in each compaction regime are schematically illustrated in Fig. 5C including: the closed-packed state where the DNA rods (depicted in blue in Fig. 5C) nearly touch each other [Fig. 5C (i)], the condensed, but not close packed, state where the DNA rods are slightly separated [Fig. 5C (ii)-a], and the expanded state where the DNA rods are well separated [Fig. 5C (iii)]. In the latter state, the charged lipids (headgroups depicted in green in Fig, 5C) tend to accumulate above/below the DNA rods. In all the other (compact) states, the DNA array is condensed by the multivalent lipids which are located primarily midway between adjacent DNA rods. Fig. 5C (ii)-a shows the stable complex associated with the "Plateau II" regime. In Fig. 5C (ii)-b, the same complex structure is in lateral phase coexistence with neutral membrane (neutral headgroups are depicted in red), just before the transition from Plateau II to the Unlocked Regime. The transition to the Unlocked Regime involves mixing of these two coexisting phases.

What remains to be discussed is the nature of the stable complex structure formed in the Plateau II region ($\Phi_1 < \Phi_{NL} < \Phi_2$). The striking feature about this structure is the fact that it is formed when the DNA rods start to move apart from each other, which may indicate that the (effective) electrostatic forces holding them together are weakened. The clue to solving this apparent contradiction comes from tracking the motion of the lipids. In the simulations, we observe that the NLs move freely within the membrane plane in the Plateau I regime, whereas the MVLs exhibit 1D diffusive motion within narrow strips located halfway between adjacent DNA rods along the DNA axis direction. Because the DNA spacing is fixed, the addition of NLs to the complex increases the lipid area density (i.e., decreases *a*, the area per lipid), which slows the motion of the lipids considerably. This trend continues even as $d_{DNA}$ increases. The increase in the total membrane area simply does not compensate for the more rapid increase in the total number of lipids. The results for the mean area per lipid are plotted against $\Phi_{NL}$ in Fig. 6A, where the solid horizontal line denotes the equilibrium area per lipid, $a^{eq} \approx 69$ Å$^2$, previously calculated for neutral membranes (28). During the transition between Plateau I and Plateau II (the different compaction regimes are indicated at the bottom of Fig. 6A), the area per lipid *a* is *smaller* than $a^{eq}$, reflecting the fact that the lipids are effectively compressed in this regime. Thus, the forces now governing the change in $d_{DNA}$ are the attractive MVL-mediated DNA–DNA electrostatic interactions and the repulsive elastic forces arising from in-plane crowding of lipids. At $\Phi_{NL} \approx \Phi_1$, the area density of the lipids exceeds the freezing density and the complexed membranes enter a gel state. In this phase, (i) the slow diffusion of both the neutral and charged lipids stops (more precisely, diffusion is not observed on the timescale of the simulations) and (ii) the lipids exhibit local (short-range) hexagonal order (see Fig. 6C). Further computational evidence for the existence of the intermediate gel phase comes from our free energy calculations and comparison to the mixed state (see Fig. 4).The lipid model that we use does not provide computational data on the ordering of the chains since these degrees of freedom are coarse-grained in our simple model where the tail is represented by two beads. However, we have taken wide angle X-ray diffraction data of samples in the gel state which only shows broad peaks. The observed lack of sharp reflections rules out a phase with chain crystallization of the lipid tails, which may be attributed to the large entropic cost associated with the freezing of the hydrocarbon chains. Ordering of headgroups cannot be detected by X-ray scattering experiments due to the low electron density contrast. The wide angle X-ray scattering data is, however, consistent with local ordering of the lipids as present in a dense liquid or an amorphous solid such as the computationally observed gel phase.

Obviously, some mixing entropy is lost due to the immobilization and local ordering of the lipids, but the gel phase appears to be thermodynamically very stable due to the optimization of the intermolecular hydrophobic interactions in the densely packed ordered membranes. In other words, the transition into a gel phase with local ordering is driven by enthalpy rather than entropy. Further addition of NLs requires breaking these hydrophobic bonds and the electrostatic bonds holding the DNA together (which disappear when the DNA spacing grows). This is avoided by the formation of a coexisting phase of neutral membranes.

**Transfection Efficiency.** The physical picture emerging from the above experimental and computational results is consistent with our TE data, which is plotted in Fig. 7 as a function of $\Phi_{DOPC}$. Based on the TE results and mechanistic studies, we have previously identified three regimes of transfection efficiency ("TE Regime I-III"): Starting at $\Phi_{DOPC}=0$, TE increases with $\Phi_{DOPC}$ in TE Regime III (extending over the Plateau I region), reaches an optimal value in TE Regime II (corresponding to the transition region between Plateau I and Plateau II), and finally exponentially decreases with $\Phi_{DOPC}$ in TE Regime I (comprising Plateau II and the Unlocked Regime), over about three orders of magnitude. A possible explanation for the initial increase in TE of tightly packed CL–DNA complexes is improved dissociation of the DNA from the oppositely charged membranes. This is required to make the DNA available for expression. Reducing the membrane charge density and thus the electrostatic attractive forces between the DNA and the cationic complex should, indeed, facilitate unbinding of the DNA rods. This trend continues until the optimal transfection efficiency is attained, which happens just before the system reaches Plateau II. Previously it has been demonstrated that the decrease in TE upon increasing $\Phi_{DOPC}$ beyond this point is due to trapping of the complexes in the endosomes (14). Transfection requires release of the complex into the cytoplasm via activated fusion with the endosomal membrane – a process which is inefficient at high $\Phi_{DOPC}$ (low $\sigma_M$). The present study points to another possible cause for the reduced TE in this regime, namely the exceptional mechanical stability exhibited by MVL-containing complexes during Plateau II. This stability hinders activated fusion and endosomal release and, therefore, would further suppress efficient transfection.

**Conclusions.** In this work, we have used a combination of experimental and computational techniques to explore and analyze the packing behavior of lamellar CL–DNA complexes containing multivalent lipids. Three distinct DNA packing regimes have been identified, corresponding to systems with low, intermediate, and large $\Phi_{NL}$, respectively. Only in the latter of these regimes, the DNA spacing increases with $\Phi_{NL}$. In the first two regimes, the DNA rods appear to be "locked", first around the minimal distance of close packing (24–26 Å for the lipid MVL5) and then around a value 2–3 Å larger than the closely-packed distance. This tight packing of the DNA rods is observed even in salt-containing DMEM at a CL/DNA charge ratio of 2.8. For comparison, the DNA spacing in complexes containing monovalent DOTAP, prepared under the same conditions, does not fall below 40 Å. (Fig. 2C). This difference clearly demonstrates the existence of attractive DNA–DNA interactions mediated by the headgroups of the multivalent lipids intercalated between the DNA rods. Thus, our work provides the first example of DNA condensation mediated by cationic lipids rather than multivalent counterions.

Quite remarkably, our simulations suggest that the less tightly-packed complexes in the Plateau II regime are more stable than the close-packed complexes in the Plateau I region. We attribute this unique stability to the appearance of a gel state of the complexed membranes, although we note that this hypothesis may be incomplete since it is based on CG simulations which ignore the fine molecular details of the system. Other mechanisms may play a role in the transition between the condensed states in Plateau I and Plateau II regimes. For instance, a switch in position of the spermine-like charged moieties on the MVL headgroups that intercalate between the DNA rods, from the major groove to the minor groove, or vice versa (29) may take place. The occurrence and possible influence of such a configurational change on the stability of the condensed states and the transition between them can only be investigated using detailed atomistic simulations. Another feature that should be inspected is the hydration of the DNA. Specifically, the slight increase in $d_{DNA}$ between the first and second plateaus may result from the formation of an additional hydration shell around the DNA rods. This may provide a complementary explanation for the thermodynamic stability of the second plateau regime.

A high degree of complex stability tends to hinder the release of the DNA to the cytoplasm and, therefore, reduce the transfection efficiency. In transfection studies of MVL-containing complexes, the maximum TE is usually achieved in the transition region between Plateau I and Plateau II, i.e., before the stable configuration of the Plateau II region is reached. The high TE in this regime reflects moderate stability of the complex, as well as the optimization of opposite effects associated with increasing $\Phi_{DOPC}$: (i) decrease in the membrane charge density vs. (ii) decrease in the electrostatic forces binding the DNA to the cationic membranes. The former slows fusion with the endosomal membrane, whereas the latter accelerates the rate of dissociation of the DNA from the complex. The structure–function relationships for CL–DNA complexes described here can serve as a guide for rational design of efficient synthetic gene therapy vectors. Our work is directly relevant to *ex vivo* gene therapy, where cells are removed, transfected in culture, and returned to patients after selection.

**Materials and Methods.**

**Synthesis, SAXS and TE Experiments.** The syntheses of MVL5 (13) and MVL3 (12) have been described. DOPC and DOTAP were obtained from Avanti Polar Lipids. Detailed procedures for MVL liposome preparation, sample preparation and data acquisition for SAXS experiments and the measurement of TE have been described previously (12,13). SAXS measurements were performed at the Stanford Synchrotron Radiation Laboratory.

**Computer Modeling and Simulations.** The details of the intermolecular short range interactions between the beads (lipid monomers) and the rods (DNA molecules) have been previously described in refs. (28) (bead-bead interactions) and (20) (bead-rod and rod-rod interactions). Other short range interactions involving the point charges are described in the main body of the paper. The long range electrostatic interactions were accounted for using the Lekner summation (20, 30), assuming that the system is at room temperature and with a bulk water uniform dielectric constant $\varepsilon = 78$ (Bjerrum length = 7.1 Å). The simulations were conducted in a rectangular cell of size of size $L_x \times L_y \times L_z$, with full periodic boundaries along the x and y directions, and periodicity with respect to only lipid mobility and short range interactions in the z direction. The rods were arranged in a one-dimensional array, parallel to the y axis and with equal spacing along the x direction. We set $L_y = 102$ Å, meaning that in the simulation cell each DNA rod, with uniform charge density $\lambda = -e/1.7$ Å, carries a total charge of $-60$ e. We study isoelectric complexes where the simulated system contains 10 equally spaced DNA rods, 120 MVLs with charge +5 (60 on each side of the DNA array), and a number of NLs corresponding to the $\Phi_{NL}$ under investigation. Monte Carlo (MC) simulations were used to generate the evolution of the lipids. Each MC step consists of (on average) an attempt to translate (and make some minute changes in the relative locations of the three beads with respect to each other) and rotate each lipid. Each MC step also included two attempts to change $L_x = 10\ d_{DNA}$, using a newly proposed sampling scheme for constant surface tension simulations (20). The simulations were performed at vanishing surface tension. The duration of the simulations for each data point was $5 \times 10^7$ MC time steps, including $5 \times 10^6$ steps for equilibration. The simulations were carried out on the "high performance on demand computing cluster" at Ben Gurion University, where the simulations of each data point run on a different cluster node.

**Acknowledgments.**
This work was supported by NIH grant GM-59288, DOE grant DE-FG0206ER46314, and NSF grant DMR-0503347. The Stanford Synchrotron Radiation Laboratory is supported by the DOE. OF thanks the staff of Ben Gurion University computation center for assisting with the simulations.

**Appendix A: list of Symbols and Abbreviations.**

| Description | Symbol |
|---|---|
| 1,2-dioleoyl-*sn*-glycerophosphatidylcholine | DOPC |
| 2,3-dioleyloxy-propyl-trimetylammonium chloride | DOTAP |
| Dulbecco's modified Eagle's medium | DMEM |
| multivalent lipids | MVLs |
| neutral lipids | NLs |
| cationic lipids | CLs |
| coarse-grained | CG |
| Monte Carlo | MC |
| small angle X-ray scattering | SAXS |
| transfection efficiency | TE |
| chemical potential of lipids in free (non-complexed) neutral membranes | $\mu_{free}$ |
| DNA interaxial distance | $d_{DNA}$ |
| energy difference per unit length of one unit cell of the complex | $\Delta$ |
| free energy | $F$ |
| free energy gain per neutral lipid added to the complex | $\mu_{complex}(\Phi_{NL})$ |
| inverted hexagonal phase of CL-DNA complexes | $H_{II}^C$ |
| lamellar phase of CL-DNA complexes | $L_\alpha^C$ |
| length of the complex parallel to the DNA axis | $L_y$ |
| length of the complex perpendicular to the DNA axis | $L_x$ |
| lipid bilayer thickness | $\delta_L$ |
| average area per lipid | $a$ |
| equilibrium area per lipid | $a^{eq}$ |
| distance from the DNA rod located at the end of the unit cell ($0 \leq x < d_{DNA}$) | $x$ |
| local area per lipid | $a(x)$ |
| local mole fraction of species i (i - NL or MVL) | $\Phi_i(x)$ |
| mean mole fraction of species i (i - NL or MVL) | $\Phi_i$ |
| membrane charge density | $\sigma_M$ |
| membrane spacing | $d$ |
| mole fraction of species i | $\Phi_i$ |
| number of NLs added to a reference state of the complex | $N$ |
| thickness of the water/DNA layer | $\delta_W$ |

**FIGURE LEGENDS**

Figure 1: Structures of the neutral lipid DOPC, the monovalent lipid DOTAP and the multivalent lipids MVL3 and MVL5 (12, 13).

Figure 2: (A) X-ray diffraction patterns of CL–DNA complexes at varied mol% of MVL3 (left) and MVL5 (right), prepared at a lipid/DNA charge ratio of 2.8 in the presence of DMEM. (B) DNA spacing as a function $\Phi_{DOPC}$ in complexes containing the MVLs. Solid symbols: systems at a charge ratio of 2.8 in DMEM; open circles: systems at a charge ratio of 1 (isoelectric point) in water. (C) DNA spacing as a function of $\Phi_{DOPC}$ in complexes containing the monovalent lipid DOTAP. Solid circles: complexes at a charge ratio of 2.8 in DMEM; open circles: complexes at a charge ratio of 1 (isoelectric point) in water; dashed line: theoretical prediction for isoelectric systems based on a simple packing argument (see refs. (18–20)).

Figure 3: Schematic depiction of tested representations of the multivalent lipid MVL5 in our CG model – a trimer consisting of three beads of diameter b, connected by stiff springs. The black and white beads represent the hydrophilic and hydrophobic parts of the lipid, respectively. The details of the intra- and intermolecular interactions between the beads appear in ref. (28). The red circle represents a +5 point charge located: (i) at the center of the hydrophilic bead, (ii) at the tip of the molecule (a distance b/2 from the center of the hydrophilic bead on the line connecting the centers of the middle and hydrophilic beads), and (iii) at the end of an inextensible tether connected to the tip of the molecule. The length of the tether $l$ is allowed to vary within the limits of $0 < l < 1.5$ b, so that the point charge may be found a distance b/2 < r < 2b from the center of the hydrophilic bead on the line connecting the centers of the middle and hydrophilic beads. The latter model (iii) has been used in the simulations reported here.

Figure 4: The excess free energy per unit length of one unit cell of the complex (see definition, Eq. 2), as a function of $\Phi_{NL}$. In the region $\Phi_1<\Phi_{NL}<\Phi_2$, the complexed membranes are in lateral phase coexistence with neutral membranes within the supramolecular assembly and, therefore, the DNA spacing remains constant.

Figure 5: Computional results: (A) DNA spacing as a function of $\Phi_{NL}$. The computational data should be compared with the experimental data for isoelectric MVL5 complexes denoted by open circles in Fig. 2B (see text). (B) Local fraction of MVLs as a function of x, the position within the unit cell of the complex. The curves correspond to the data points plotted in (A) as solid black, red and blue circles, respectively. (C) Schematics illustrating the different structures formed in each compaction regime. DNA rods are shown in blue, multivalent headgroups in green and neutral headgroups in red. The depicted structures include: (i) the close-packed state where the DNA rods nearly touch each other, (ii) the condensed, but not close-packed, state where the DNA rods are slightly separated, and (iii) the expanded state where the DNA rods are well separated. Fig. 5C (ii)-a shows the stable complex phase associated with the "Plateau II" regime. In Fig. 5C (ii)-b, the same complex structure appears in lateral phase coexistence with neutral membrane, just before the transition from Plateau II to the Unlocked Regime.

Figure 6: (A) The area per lipid as a function of $\Phi_{NL}$. The regions corresponding to distinct DNA compaction regimes are indicated at the bottom of the figure: (from left to right) Plateau I, transition from Plateau I to Plateau II, Plateau II (which is a region of lateral phase coexistence), and the Unlocked Regime. During the transition between Plateau I and II, the area per lipid drops below the equilibrium area per lipid in neutral membranes (indicated by the solid horizontal line) and the fluid complex membranes transition into a gel state. (B–D) Snapshots of simulations showing typical equilibrium configurations of the complex in Plateau I (B), Plateau II (C), and the Unlocked Regime (D). In the snapshots, the lipids located on one side of the DNA array are shown, where each lipid is represented by the position of its middle bead (see Fig. 3). The beads are depicted in grey and DNA rods in red. In (B) and (D) the complex membranes are fluid, while in (C) they are in the gel state. The membranes in the gel state exhibit local ordering of the lipids which do not diffuse within the membrane plane.

Figure 7: Transfection Efficiency of MVL/DOPC–DNA complexes plotted as a function of $\Phi_{DOPC}$ for complexes containing the multivalent lipids MVL3 (red squares) and MVL5 (blue circles). Note three regimes of transfection efficiency ("TE Regime I-III") as a function of $\Phi_{DOPC}$ (12) and their relation to the compaction regimes: Starting at $\Phi_{DOPC}=0$, TE increases with $\Phi_{DOPC}$ in TE Regime III (corresponding to Plateau I) through an optimum in TE Regime II (corresponding to the transition region between Plateau I and Plateau II). With further increase of $\Phi_{DOPC}$ in TE Regime I (comprising Plateau II and the Unlocked Regime), TE decreases exponentially over about three orders of magnitude.

**FIGURE 1**

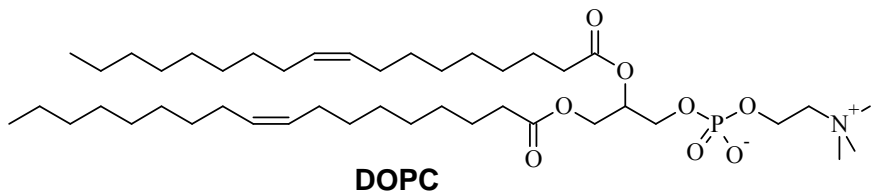
**DOPC**

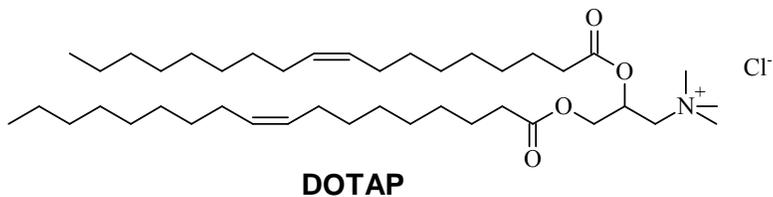
**DOTAP**

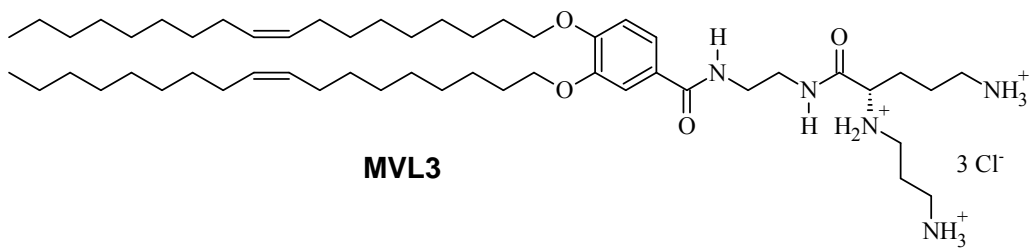
**MVL3**

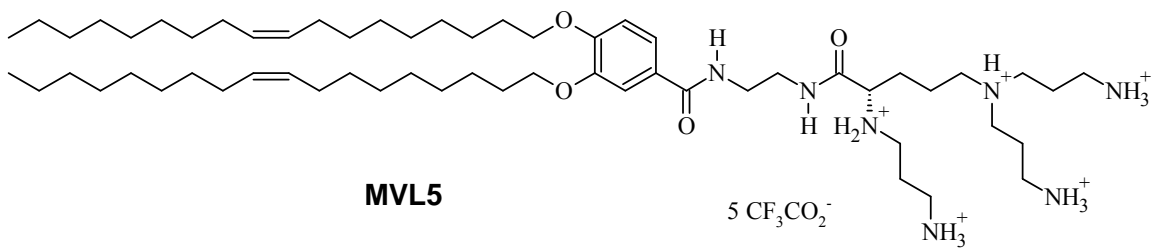
**MVL5**

**FIGURE 2**

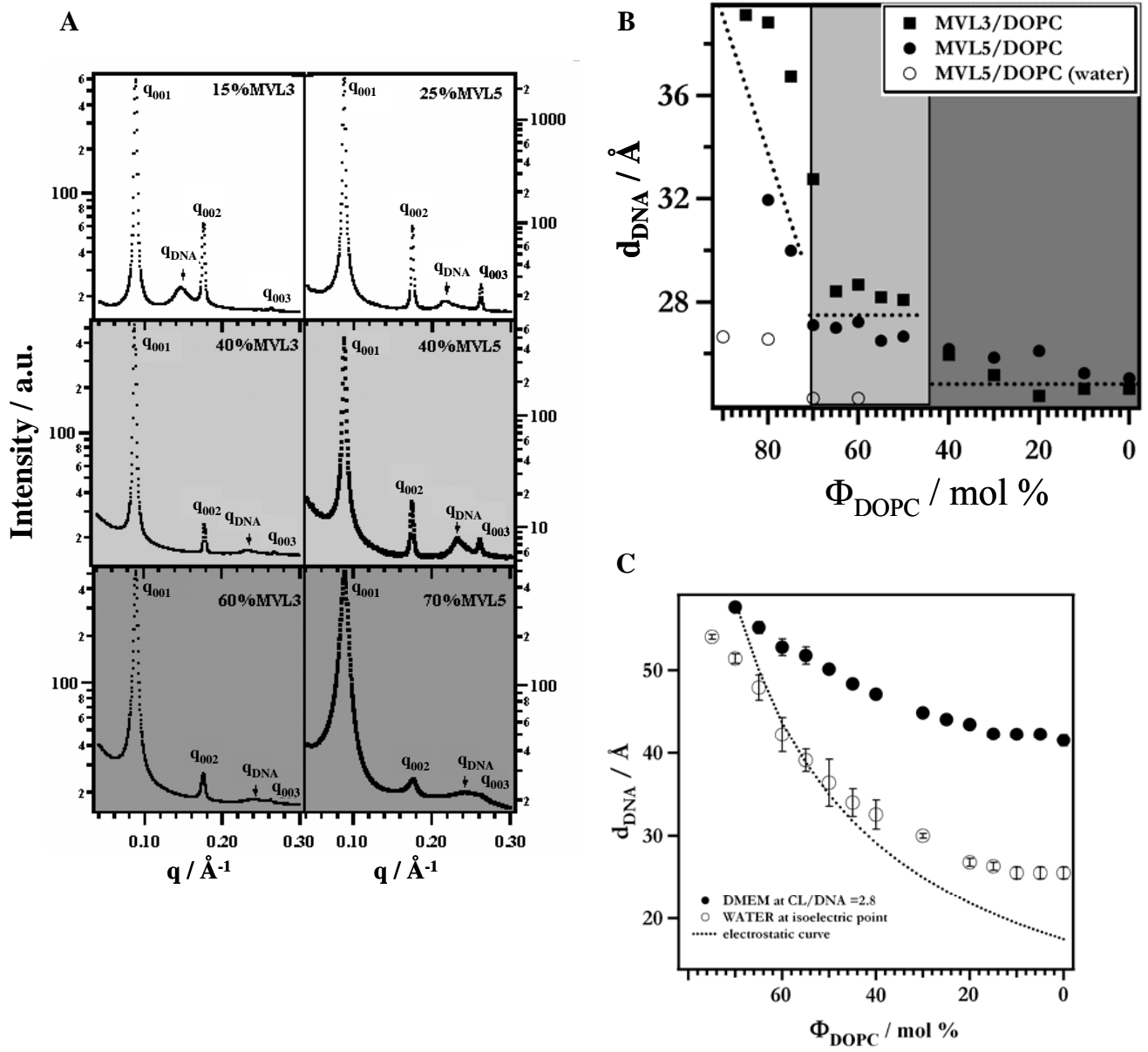

18**FIGURE 2**

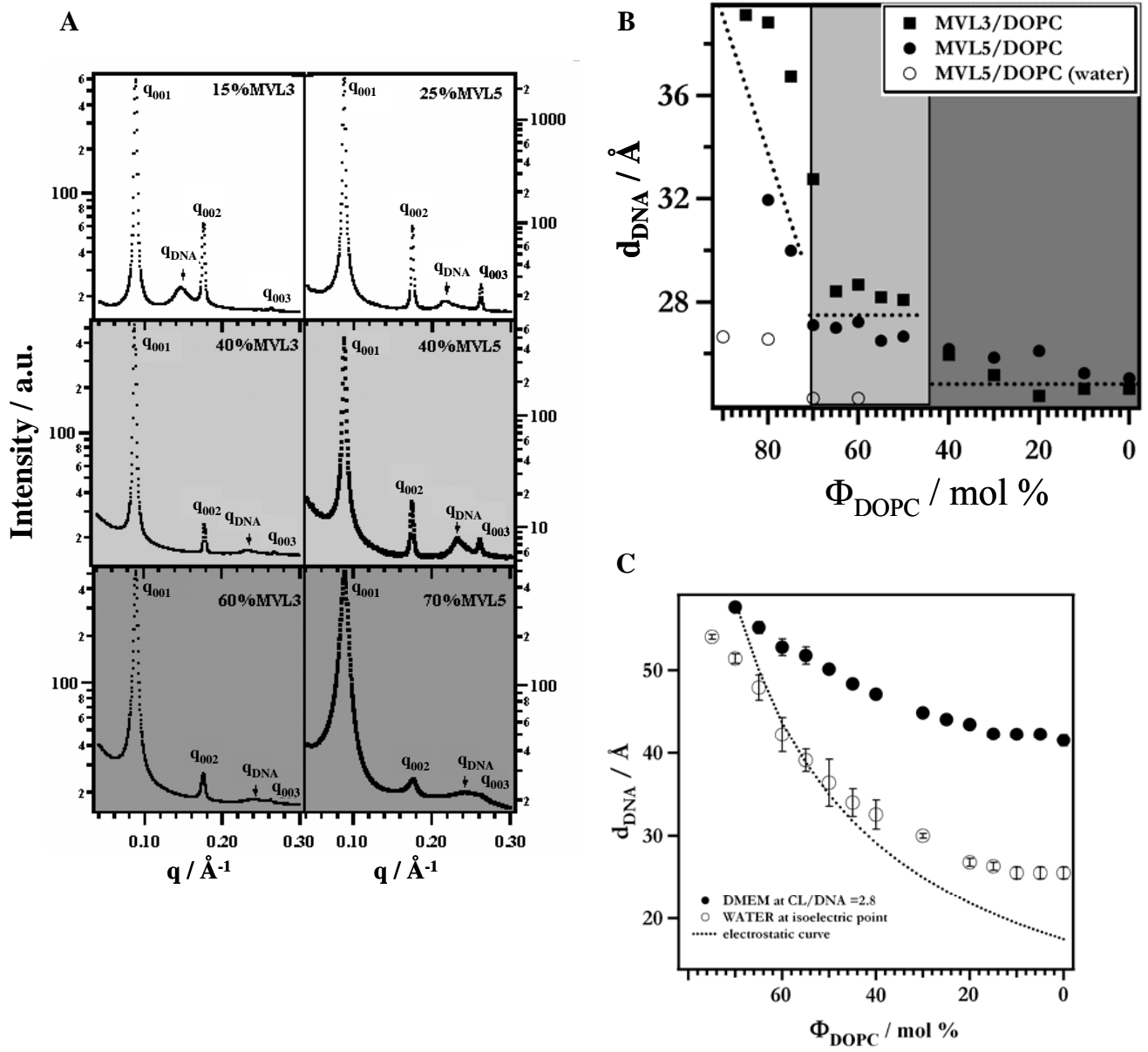

**FIGURE 3**

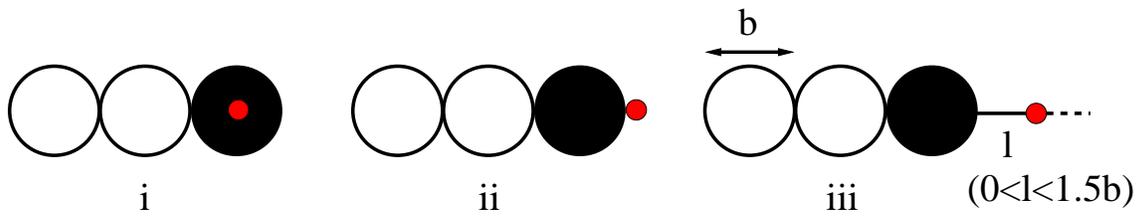

**FIGURE 4**

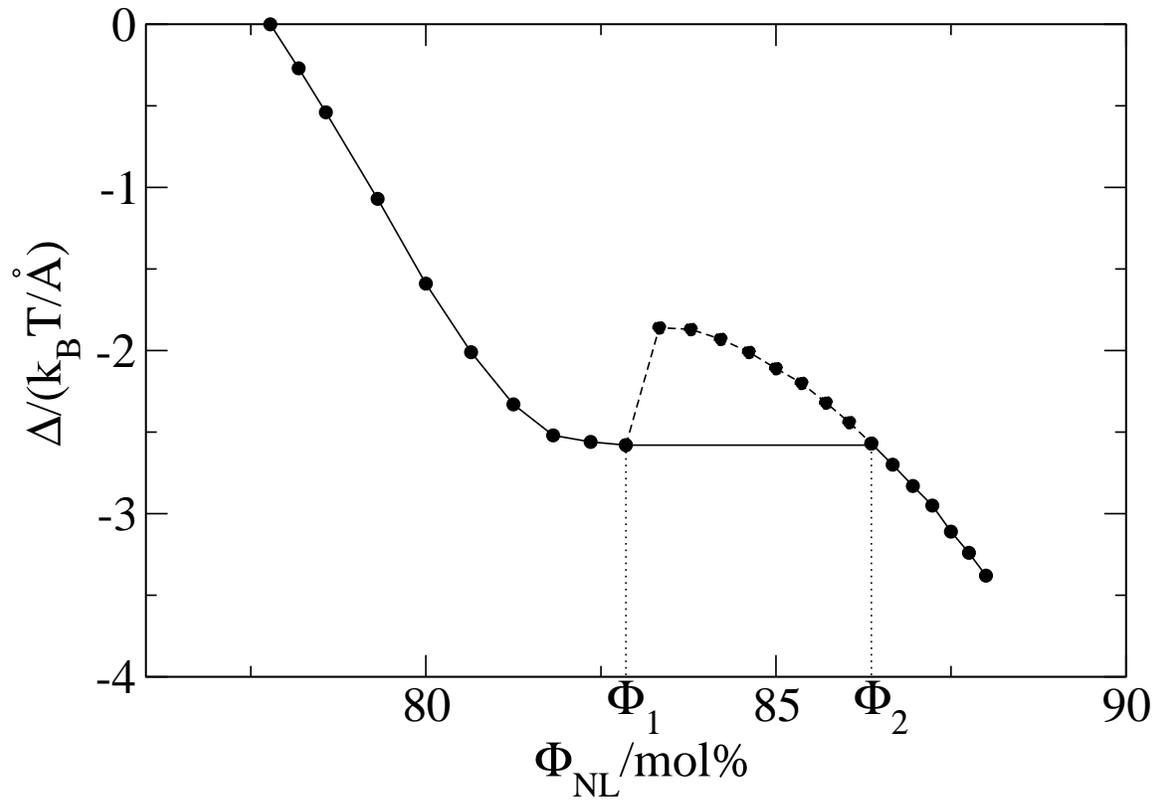

**FIGURE 5**

A

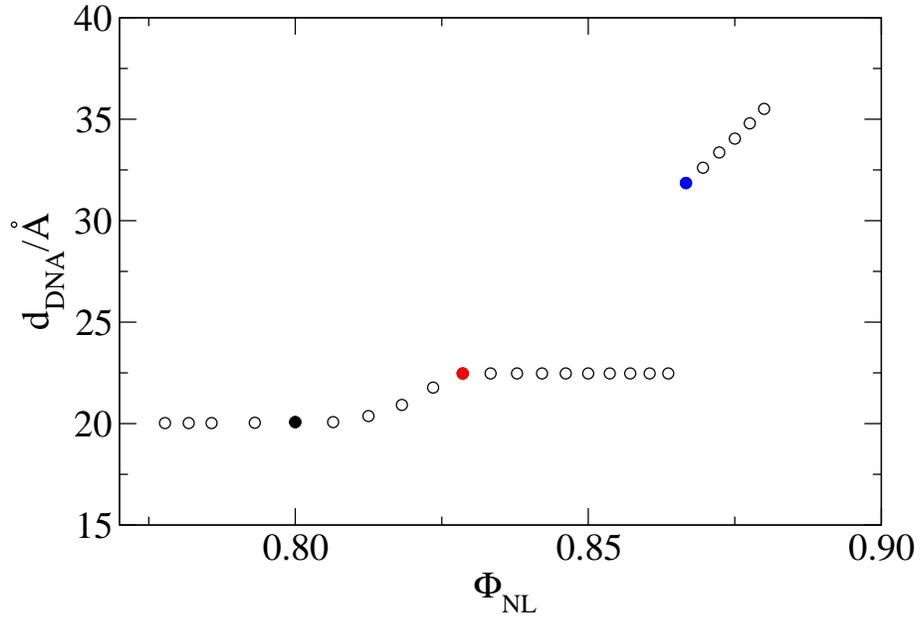

B

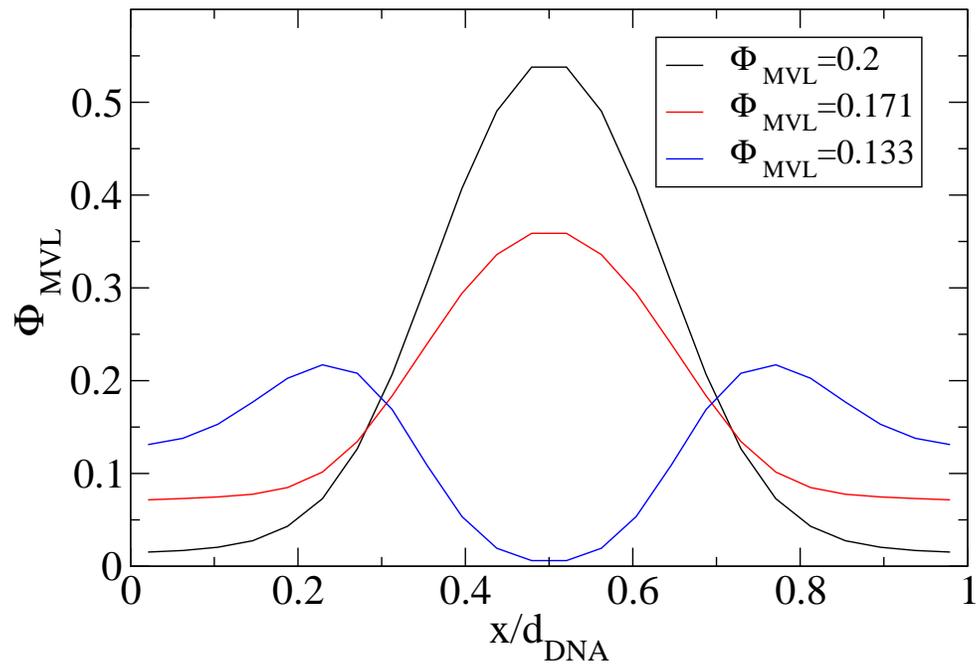

**FIGURE 5 (cont)**

**C**

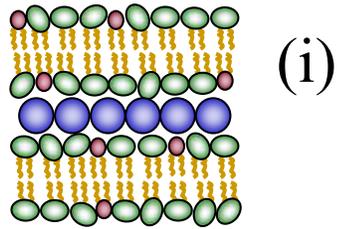 (i)

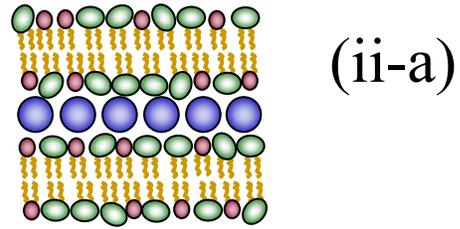 (ii-a)

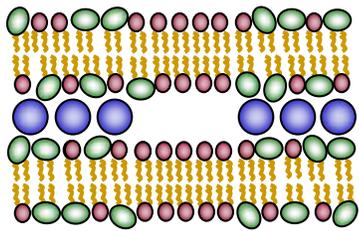
(ii-b)

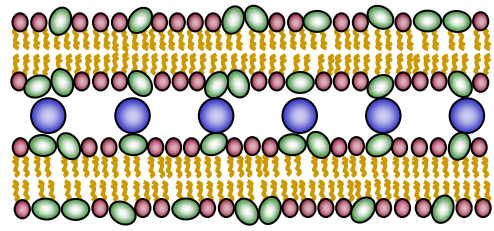
(iii)

**FIGURE 6**

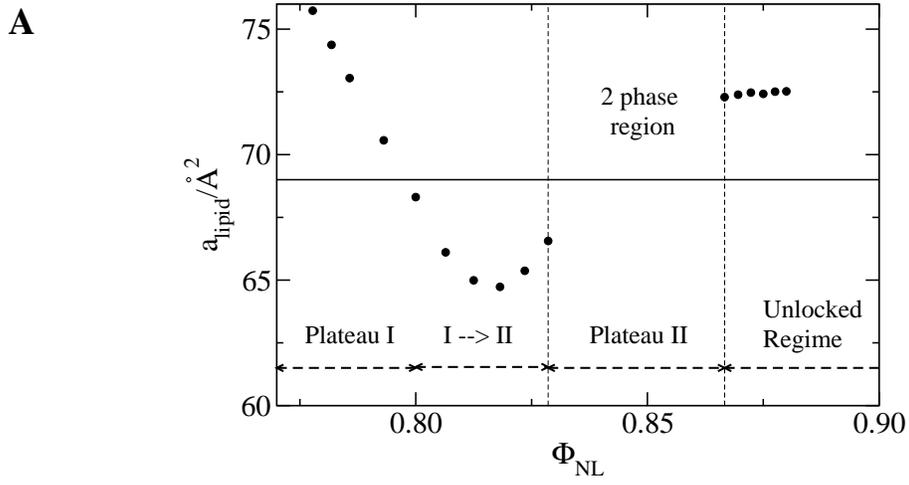

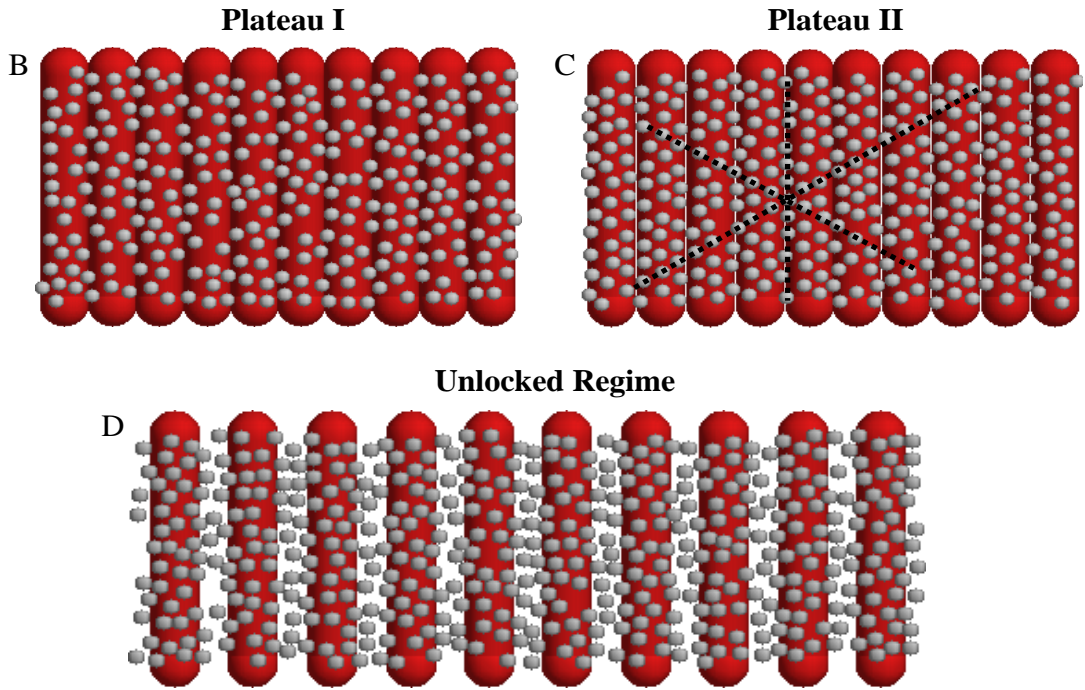

**FIGURE 7**

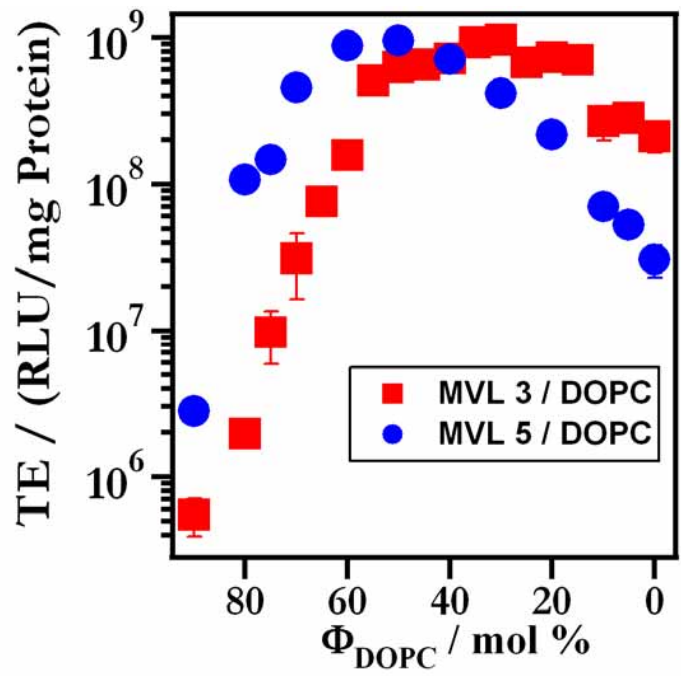